\documentclass[journal]{IEEEtran}

\usepackage[utf8]{inputenc}
\usepackage{cite}
\usepackage{amsmath,amssymb,amsfonts}
\usepackage{graphicx}
\usepackage{booktabs}
\usepackage{multirow}
\usepackage{array}
\usepackage{tabularx}
\usepackage{makecell}
\usepackage{url}
\usepackage{algorithm}
\usepackage{algorithmic}
\usepackage{siunitx}
\usepackage[hidelinks,colorlinks=true,linkcolor=blue,citecolor=blue,urlcolor=blue]{hyperref}
\usepackage[table]{xcolor}
\usepackage{balance}
\usepackage{stfloats}
\usepackage{threeparttable}

\newcolumntype{Y}{>{\raggedright\arraybackslash}X}
\newcolumntype{C}{>{\centering\arraybackslash}X}
\title{ATCCaps: A Call-Sign-Aware Speech Dataset for Air Traffic Control Recognition}

\author{Dongdong~Li$^*$, Jianwei~Song, Jianwei~Wang, and Zhe~Wang
\thanks{The authors are with the Department of Computer Science and Technology, East China University of Science and Technology, 130 Meilong Road, Shanghai 200237, China (e-mail: ldd@ecust.edu.cn; wangzhe@ecust.edu.cn).}%
\thanks{Manuscript received February 7, 2025; This work was supported in part by the Natural Science Foundation of China under Grant 62276098. (Corresponding author: DongDong Li.)}}

\begin{document}
\maketitle

\begin{abstract}
Call signs are safety-critical entities in air traffic control (ATC)
communications because they identify the target aircraft of each spoken
instruction. This paper presents ATCCaps, a call-sign-aware ATC speech
dataset with caption-level audio-text supervision. Built from real ATC
radiotelephony recordings, ATCCaps contains 202.94 hours of curated
audio, 170,385 utterances, and 922 unique normalized call signs. The
construction pipeline combines confidence-aware transcript parsing,
ADS-B-derived call-sign metadata, call-sign normalization, rule-based
quality filtering, and LLM-assisted caption generation. Each retained
sample is paired with transcript descriptions, call-sign descriptions,
and ATC-style captions, supporting ASR evaluation, call-sign matching,
and call-sign-aware audio-text retrieval. We further characterize ATCCaps through split statistics, call-sign
coverage, seen/unseen call-sign analysis, filtering audits, and caption
quality evaluation. The evaluation subset is derived from the
human-annotated ATCO2-test-set, enabling reference evaluation with
manual transcripts. Results show that ATCCaps provides scalable
audio-grounded call-sign supervision, while caption analysis highlights
the need to explicitly validate call-sign and numeric fidelity. Reference
ASR and CLAP-based baselines demonstrate the usability of ATCCaps for
call-sign-aware ATC speech modeling.
\end{abstract}

\begin{IEEEkeywords}
Air traffic control, call-sign recognition, speech dataset, audio-text grounding, caption-level supervision
\end{IEEEkeywords}

\section{Introduction}
Air traffic control (ATC) communication remains a safety-critical
radiotelephony process between air traffic controllers and pilots.
Lin~\cite{lin2021spoken} reviewed spoken instruction understanding in
ATC and emphasized its importance for safety monitoring, controller
assistance, and intelligent air traffic operation. Badrinath and
Balakrishnan~\cite{badrinath2022automatic} further showed that ASR can
support practical ATC communication analysis, while also revealing the
difficulty of recognizing operationally important words under real radio
conditions. Recent ATC-ASR studies further highlight domain-specific
challenges, including fast speech, volatile background noise, limited
annotated resources, accented communication, and the frequent appearance
of named entities~\cite{fan2023speech,wee2025adapting}. These studies
indicate that ATC speech
recognition requires more than general word-level transcription: the
recognized output must preserve the operational entities that determine
the meaning and safety impact of an instruction.

Progress in general speech and audio representation learning has been
driven by self-supervised speech encoders, weakly supervised ASR, and
audio-text pretraining~\cite{baevski2020wav2vec,hsu2021hubert,radford2023robust,zheng2021fused,zhang2024speechlm,wu2023large}.
However, these general-purpose advances do not by themselves provide the
normalized call-sign supervision needed for ATC entity grounding.

Among ATC-related entities, call signs are particularly important because
they identify the target aircraft of a spoken instruction. Pellegrini
et al.~\cite{pellegrini2018airbus} organized a call-sign identification
challenge and showed that extracting airline call signs from ATC speech
is a dedicated and nontrivial task. Juan et al.~\cite{zuluaga2020automatic}
improved airline call-sign identification by jointly exploiting radar
trajectories and audio information, demonstrating the value of external
traffic context. Badrinath and Balakrishnan~\cite{badrinath2022automatic}
studied the extraction of callsigns and runway numbers from real ATC
communications, further confirming that safety-critical entities deserve
separate evaluation. Their analysis also points out that
digit-composed entities, such as call signs, frequencies, flight levels,
and runway numbers, are especially vulnerable to acoustic confusion and
long-tail vocabulary effects.

Contextual ASR studies provide another line of evidence for the central
role of call signs. Oualil et al.~\cite{oualil2015dynamic} introduced
dynamic context information into real-time ATC speech recognition,
showing that operational context can guide recognition. Kocour et
al.~\cite{blatt2022call} boosted contextual information in ASR for
air-traffic call-sign recognition by incorporating call-sign knowledge
during decoding. Nigmatulina et al.~\cite{nigmatulina2021improving} proposed a
two-step strategy that leverages contextual data for ATC speech
recognition, where surveillance-derived information assists recognition
and named-entity correction. Kasttet et al.~\cite{kasttet2023toward}
used fuzzy string matching between ASR hypotheses and ADS-B data for
aircraft call-sign detection, illustrating the practical relevance of
surveillance-linked call-sign supervision. Guo et al.~\cite{badrinath2022automatic}
systematically compared contextual ASR approaches and formulated
dynamic contextual knowledge as call signs extracted from surveillance
radar and flight-plan information. These works motivate the construction
of datasets that expose call-sign supervision, call-sign candidates, and
audio-grounded call-sign descriptions in a reusable form.

Existing ATC data resources provide complementary foundations for
speech recognition and spoken instruction understanding. ATCOSIM
~\cite{hofbauer2008atcosim} provides clean simulated ATC speech with
transcripts and speaker metadata, supporting controlled ASR studies.
ATCSpeech~\cite{yang2019atcspeech} moves toward real operational
pilot-controller communication with manual transcription, while
ATCO2~\cite{zuluaga2022atco2,gomez2024atco2} substantially increases the scale and
diversity of public ATC speech resources by collecting large amounts of
real-world radiotelephony data. These corpora have advanced ATC-ASR
research, but their annotation schemes mainly focus on transcription or
metadata rather than normalized call-sign grounding and audio-aligned
caption descriptions.

Semantic annotation resources have further promoted ATC spoken
instruction understanding. Oualil et al.~\cite{oualil2015dynamic}
introduced semantic concepts and dynamic context into ATC speech
recognition. Lin~\cite{lin2021spoken} reviewed ATC spoken instruction
understanding and summarized its main challenges, techniques, and
applications. Badrinath and Balakrishnan~\cite{badrinath2022automatic}
focused on extracting operational entities such as callsigns and runway
numbers from real ATC communications. Guo et al.~\cite{guo2023m2ats}
integrated ATC voice, instruction text, and trajectory information for
multimodal ATC tasks, and Zhang et al.~\cite{zhang2025atsiu}
developed a professional intent-slot benchmark for text-level ATC
spoken instruction understanding. These studies enrich semantic
supervision for ATC text and selected operational entities, while
large-scale audio-grounded call-sign supervision remains insufficiently
explored.

At the same time, prior work on speech retrieval, contrastive
audio-text learning, and multimodal representation learning has shown
that paired audio-text descriptions can improve retrieval and transfer
learning~\cite{audhkhasi2017end,chung2020perfect,lou2022audio,le2020contrastive,zhang2022contrastive,hu2024comprehensive}.
ATCCaps adapts this supervision idea to ATC by making the textual side
explicitly call-sign-aware.

Motivated by this gap, we present ATCCaps, a call-sign-aware
ATC speech dataset with normalized call-sign supervision and
caption-level audio-text grounding. ATCCaps is constructed from ATCO2
through confidence-aware transcript parsing, ADS-B-guided call-sign
candidate extraction, call-sign normalization, rule-based quality
filtering, and LLM-assisted caption generation. The resulting dataset
contains 202.94 hours of curated ATC audio, 170,385 utterances, and
922 unique call signs. Each retained utterance is paired with audio,
transcript metadata, normalized call-sign information, and
call-sign-aware textual descriptions, supporting ASR evaluation,
call-sign recognition, call-sign-aware retrieval, and seen/unseen
call-sign generalization.

The main contributions of this work are summarized as follows:
\begin{itemize}
    \item We construct ATCCaps, a call-sign-aware ATC speech dataset
    containing 202.94 hours of curated audio, 170,385 utterances, and
    922 unique normalized call signs, together with paired audio,
    transcripts, call-sign metadata, and caption-level descriptions.

    \item We design a reproducible construction pipeline that combines
    confidence-aware transcript parsing, ADS-B-guided call-sign candidate
    extraction, call-sign normalization, and rule-based filtering according
    to signal quality, duration, and call-sign validity.

    \item We introduce LLM-assisted call-sign caption annotation to generate
    audio-aligned textual descriptions for call-sign grounding and
    retrieval. The augmentation expands the textual supervision of the
    training split and records explicit call-sign corrections for quality
    analysis.

    \item We provide comprehensive dataset characterization, including
    split-level statistics, call-sign overlap analysis, seen/unseen
    call-sign settings, comparison with representative ATC datasets,
    caption-quality validation, and reference baselines for ASR,
    call-sign matching, and call-sign-aware audio-text retrieval.
\end{itemize}

The remainder of this paper is organized as follows.
Section~\ref{sec:related_work} reviews ATC datasets and call-sign
recognition studies. Section~\ref{sec:construction} describes the
ATCCaps construction pipeline. Section~\ref{sec:analysis} presents
dataset statistics, comparison, and quality validation.
Section~\ref{sec:benchmarks} reports benchmark protocols and reference
results. Section~\ref{sec:conclusion} concludes the paper.

\section{Related Work}
\label{sec:related_work}

\subsection{ATC Speech and Spoken-Instruction Datasets}
\label{subsec:datasets}

Existing ATC data resources can be broadly divided into speech corpora
for ASR and semantic datasets for spoken instruction understanding.
ATCOSIM~\cite{hofbauer2008atcosim} provides clean simulated ATC speech
with manual transcripts, making it suitable for controlled ASR studies.
ATCSpeech~\cite{yang2019atcspeech} moves toward real operational
communications by collecting multilingual pilot-controller speech with
manual transcription. ATCO2~\cite{zuluaga2022atco2,gomez2024atco2} further increases
the scale and acoustic diversity of public ATC speech resources through
large-scale radiotelephony recordings. These corpora have promoted
ATC-ASR research by providing increasingly realistic audio conditions,
but their annotations are mainly transcript-oriented and provide limited
support for structured call-sign grounding.

Semantic ATC datasets have enriched the supervision of spoken
instructions. Oualil et al.~\cite{oualil2015dynamic} introduced semantic
concepts and dynamic context for ATC speech recognition. Lin et
al.~\cite{lin2021spoken} reviewed ATC spoken instruction understanding
as a safety-critical research direction. Badrinath and
Balakrishnan~\cite{badrinath2022automatic} focused on extracting
operational entities such as callsigns and runway numbers from real ATC
communications. More recent resources, including M2ATS~\cite{guo2023m2ats}
and ATSIU~\cite{zhang2025atsiu}, further extend ATC semantic modeling
with multimodal information or professional intent-slot annotations.
These datasets are valuable for text-level semantic parsing, intent
classification, and slot filling. However, existing resources still offer
limited large-scale supervision that directly aligns audio segments with
normalized call signs, call-sign candidates, and caption-style textual
descriptions.

\subsection{Call-Sign Recognition and Contextual ASR}
\label{subsec:callsign_context}

Call-sign recognition has been widely studied because the call sign
identifies the target aircraft in ATC communication. Pellegrini et
al.~\cite{pellegrini2018airbus} treated call-sign identification as a
dedicated challenge, showing that the task remains difficult under real
ATC speech conditions. Juan et al.~\cite{zuluaga2020automatic} improved
airline call-sign identification by combining radar trajectories with
audio information. Other studies used grammar rules, ASR hypotheses, or
surveillance data to detect and correct callsigns in operational
communications~\cite{badrinath2022automatic,zuluaga2020automatic}.
These works show that call signs require entity-level treatment rather
than evaluation only through global ASR metrics.

Related aviation and ATC communication studies also emphasize that
operational context, channel characteristics, and communication
infrastructure affect speech understanding and safety-critical entity
recognition~\cite{amin2023low,weber2024air,chen2023effects,li2022recent}.

Contextual ASR provides an effective way to improve recognition of
ATC-related entities. Dynamic or static context can be injected through
language-model biasing, rescoring, fuzzy matching, or neural contextual
fusion~\cite{lin2021spoken,blatt2022call,nigmatulina2021improving,kasttet2023toward}.
Recent comparative work further shows that contextual knowledge is
particularly useful for recognizing named entities such as call signs,
runways, waypoints, and flight levels, and proposes entity-level metrics
including CSA, IPA, and IRA for ATC-ASR evaluation~\cite{badrinath2022automatic}.
Most existing studies focus on model-side use of contextual information.
Reusable data resources that provide aligned audio, normalized call
signs, call-sign candidate context, and call-sign-oriented text
descriptions remain comparatively limited.

Table~\ref{tab:dataset_comparison} summarizes representative ATC
speech and spoken-instruction datasets from the perspectives of modality,
scale, call-sign supervision, intent-slot annotation, contextual
information, and caption availability. As shown in
Table~\ref{tab:dataset_comparison}, existing resources have advanced
ATC-ASR, intent detection, slot filling, and multimodal trajectory-related modeling. However, large-scale audio-grounded call-sign supervision with caption-level textual descriptions remains limited.

\begin{table*}[!t]
\centering
\caption{Comparison of representative ATC speech, spoken-instruction, and call-sign-related datasets.}
\label{tab:dataset_comparison}
\small
\setlength{\tabcolsep}{4.2pt}
\renewcommand{\arraystretch}{1.12}
\begin{tabular}{
    >{\centering\arraybackslash}m{2.3cm}
    >{\centering\arraybackslash}m{2.2cm}
    >{\centering\arraybackslash}m{1.7cm}
    >{\centering\arraybackslash}m{1.2cm}
    >{\centering\arraybackslash}m{1.4cm}
    >{\centering\arraybackslash}m{1.1cm}
    >{\centering\arraybackslash}m{1.5cm}
    >{\centering\arraybackslash}m{1.3cm}
}
\toprule
\textbf{Dataset} &
\textbf{Modality} &
\textbf{Scale} &
\textbf{CS Sup.} &
\textbf{Intent/Slot} &
\textbf{Cap.} &
\textbf{Context} &
\textbf{Avail.} \\
\midrule

ATCOSIM~\cite{hofbauer2008atcosim} &
Aud. + Trans. &
10.7h / 10k &
No &
No / No &
No &
No &
Public \\

ATCSpeech~\cite{yang2019atcspeech} &
Aud. + Trans. &
59h / 22.7k &
Limited &
No / No &
No &
No &
Non-comm. \\

UWB-ATCC~\cite{smidl2019atcc} &
Aud. + Trans. &
20h / -- &
Limited &
No / No &
No &
No &
Public \\

ATCO2~\cite{gomez2024atco2} &
Aud. + Trans. + Meta &
5281h / -- &
Weak &
Limited &
No &
Meta &
1h free / ELDA \\

Oualil et al.~\cite{oualil2015dynamic} &
Aud. / Text &
3h / -- &
Partial &
No / 3 &
No &
Dynamic &
Private \\

Badrinath et al.~\cite{badrinath2022automatic} &
Aud. / Text &
140h / 28k &
Yes &
No / 2 &
No &
No &
Private \\

M2ATS~\cite{guo2023m2ats} &
Aud. + Text + Traj. &
104h / 110k+ &
Yes &
16 / 51 &
No &
Traj. &
Private \\

ATSIU~\cite{zhang2025atsiu} &
Text from ATC audio &
206h / 19.8k &
As slot &
9/26 / 78 &
No &
No &
Public \\

\textbf{ATCCaps} &
\textbf{Aud. + Trans. + Cap.} &
\textbf{202.94h / 170.4k} &
\textbf{Yes} &
\textbf{No / No} &
\textbf{Yes} &
\textbf{ADS-B CS} &
\textbf{This work} \\

\bottomrule
\end{tabular}
\vspace{2pt}
\begin{flushleft}
\footnotesize
\textit{Notes:} Aud. = audio; Trans. = transcript; Meta = metadata;
Traj. = trajectory; Cap. = caption-level textual description;
CS Sup. = call-sign supervision. ``As slot'' indicates that call-sign
information is included as part of slot annotations rather than provided
as normalized audio-grounded call-sign supervision. ``--'' denotes
information not explicitly reported or not directly comparable.
\end{flushleft}
\end{table*}

\section{ATCCaps Construction}
\label{sec:construction}

\begin{figure*}[t]
    \centering
    \includegraphics[width=0.95\textwidth]{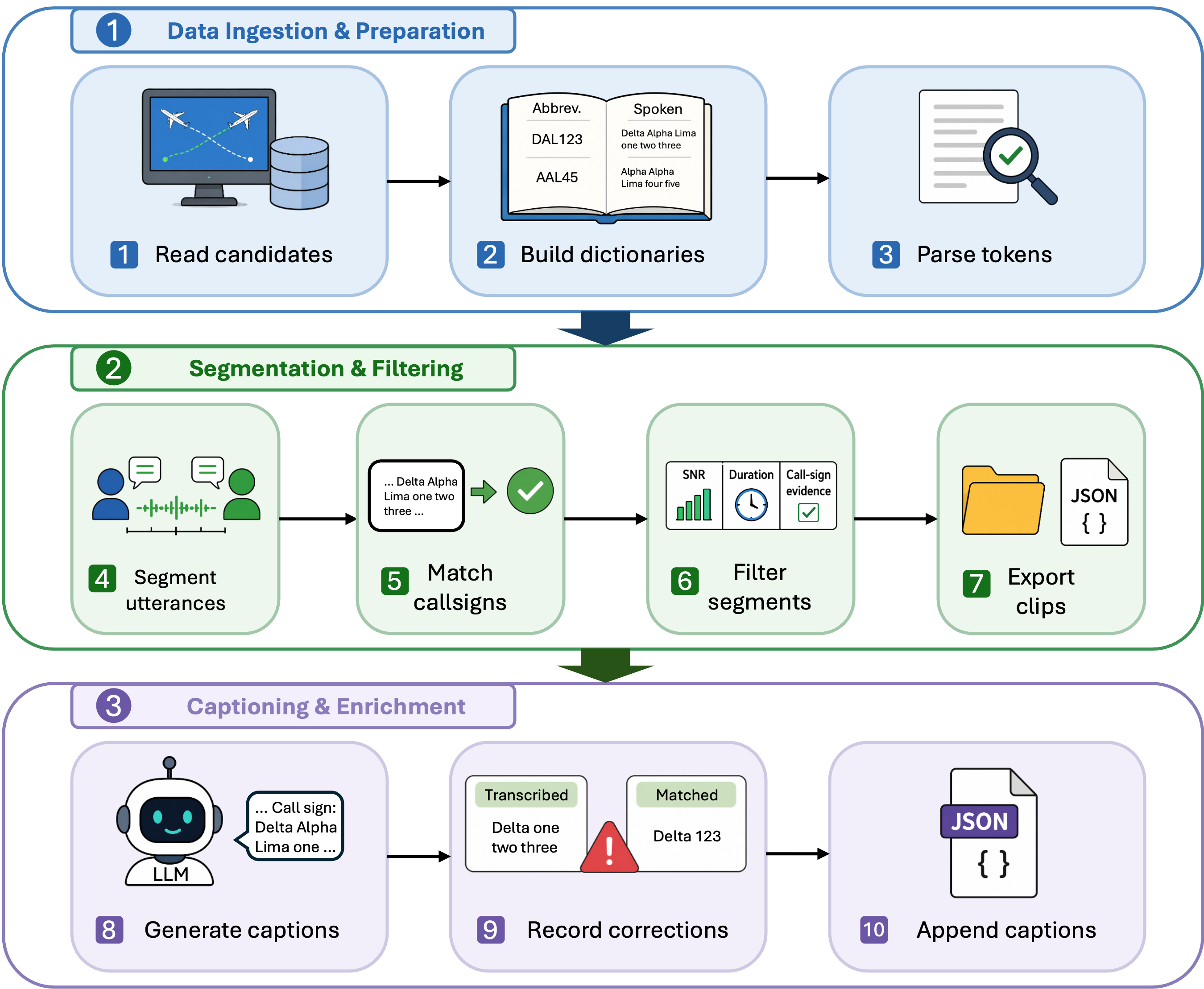}
    \caption{Overview of the ATCCaps construction pipeline. The
    ATCO2-PL branch parses machine-transcription files into speaker-turn
    utterances and uses ADS-B-derived flight-aid metadata from ATCO2
    metadata files to construct normalized call-sign dictionaries.
    Training and validation samples are filtered by SNR, duration, and
    valid call-sign evidence. The ATCO2-test-set branch uses
    human-annotated transcripts and applies call-sign matching without
    the ATCO2-PL duration threshold. Retained clips are exported as FLAC
    files with JSON text metadata, and LLM-assisted augmentation appends
    call-sign-aware ATC-style captions.}
    \label{fig:pipeline}
\end{figure*}

This section describes the construction pipeline of ATCCaps. As shown in
Fig.~\ref{fig:pipeline}, the pipeline consists of two source branches and
one annotation-enrichment branch. The first source branch processes
ATCO2-PL-set, from which the training and validation splits are built.
In this branch, machine transcriptions are parsed into speaker-turn
utterances, ADS-B-derived flight-aid metadata is used to construct
call-sign dictionaries, and the resulting segments are filtered by SNR,
duration, and valid call-sign evidence. The second source branch
processes the human-annotated ATCO2-test-set, from which the test split
is exported after call-sign matching. Finally, retained clips are paired
with JSON text metadata and enriched with LLM-assisted ATC-style
captions.

\subsection{Corpus Selection and Split Design}
\label{subsec:source_split}

ATCCaps is constructed from ATCO2~\cite{gomez2024atco2}, a large-scale
corpus of real ATC radiotelephony recordings. We use ATCO2-PL-set as
the source of the training and validation data because it provides a
large amount of realistic ATC speech covering diverse airports, accents,
radio channels, and communication scenarios. Although its transcriptions
are pseudo labels and contain noise, the scale and diversity of
ATCO2-PL-set make it suitable for constructing a call-sign-aware
audio-text resource after additional parsing and filtering.

The held-out test data are derived from the human-annotated
ATCO2-test-set. This design allows the final evaluation to rely on
manually verified transcripts rather than only pseudo-transcribed
training data. The training and validation splits follow a month-ordered
partition of ATCO2-PL-set. Specifically, monthly transcription files are
sorted chronologically; the first 13 monthly files, from 2020-10 to
2021-10, are assigned to the training split, and the last three monthly
files, from 2021-11 to 2022-01, are assigned to the validation split.
The test split is constructed separately from
ATCO2-test-set and is not sampled from the ATCO2-PL train/validation
months. Table~\ref{tab:split_rules} summarizes the split construction
rules.

\begin{table*}[t]
\centering
\caption{Split construction rules for ATCCaps.}
\label{tab:split_rules}
\small
\setlength{\tabcolsep}{4pt}
\renewcommand{\arraystretch}{1.15}
\begin{tabular}{
    >{\centering\arraybackslash}m{1.7cm}
    >{\centering\arraybackslash}m{2.2cm}
    >{\centering\arraybackslash}m{2.5cm}
    >{\centering\arraybackslash}m{3.3cm}
    >{\centering\arraybackslash}m{3.9cm}
}
\toprule
\textbf{Split} &
\textbf{Source} &
\textbf{Segmentation} &
\textbf{Selection rule} &
\textbf{Filtering rule} \\
\midrule

Train &
ATCO2-PL-set &
Speaker-turn segments &
First 13 months (2020-10--2021-10) &
SNR $>10$, duration $\geq 1.0$ s, and valid call-sign evidence \\

Validation &
ATCO2-PL-set &
Speaker-turn segments &
Last 3 months (2021-11--2022-01) &
SNR $>10$, duration $\geq 1.0$ s, and valid call-sign evidence \\

Test &
ATCO2-test-set &
Human-annotated clips &
Filtered human-transcribed subset &
Valid call-sign evidence only; ATCO2-PL duration threshold is not applied \\

\bottomrule
\end{tabular}
\end{table*}

\subsection{Preprocessing Pipeline}
\label{subsec:preprocessing}

The preprocessing stage converts ATCO2 recordings and metadata into
call-sign-bearing utterances. It contains three main operations:
call-sign dictionary construction, machine-transcription parsing, and
quality filtering. Algorithm~\ref{alg:atccaps_pipeline} summarizes the
overall procedure.

\begin{algorithm}[t]
\caption{ATCCaps Construction Pipeline}
\label{alg:atccaps_pipeline}
\footnotesize
\begin{algorithmic}[1]
\REQUIRE ATCO2 machine-transcription files; ADS-B-derived flight-aid metadata; supplementary call-sign list; LLM API
\ENSURE Filtered FLAC audio clips and JSON text metadata
\STATE Read candidate call signs from ATCO2 flight-aid metadata
\STATE Construct call-sign dictionaries from abbreviated and spoken forms
\STATE Parse machine transcriptions by selecting highest-confidence tokens
\STATE Segment parsed tokens into utterances by speaker turns
\STATE Match decoded transcripts with normalized valid call signs
\STATE Filter ATCO2-PL segments by SNR, duration, and call-sign evidence
\STATE Export retained audio clips and initial JSON text metadata
\STATE Generate ATC-style captions with the LLM
\STATE Record call-sign correction information when mismatches are detected
\STATE Append generated captions to the JSON \texttt{text} list
\end{algorithmic}
\end{algorithm}

\subsubsection{Call-sign dictionary construction}

Candidate call signs are obtained from ADS-B-derived flight-aid metadata
distributed with ATCO2. In practice, ATCO2 metadata provides a
recording-level candidate set of nearby aircraft call signs. This
metadata has already been associated with the corresponding recording
context by the ATCO2 data provider. Therefore, ATCCaps does not perform
additional ADS-B trajectory interpolation or define a new second-level
audio--ADS-B matching window. The metadata is used to construct a
candidate call-sign dictionary, and the final utterance-level decision
is made by matching the decoded transcript.

The call-sign dictionary is built from abbreviated call-sign codes and
their spoken full forms. Abbreviated call signs are detected using the
regular expression \texttt{[A-Z]\{2,3\}\textbackslash d\{1,4\}}, which
corresponds to two or three uppercase letters followed by one to four
digits. Each detected abbreviation is mapped to the corresponding spoken
full form when available, and the spoken form is converted to lowercase
for consistent text matching. For the ATCO2-test-set, a supplementary
manually curated call-sign list is used to cover spoken call signs that
are present in the human-annotated test data but not fully covered by
the original metadata. In the ATCO2-PL explicit-field pipeline,
abbreviated codes are required to contain at least three digits before
entering the valid dictionary, which reduces false matches caused by
short or unstable abbreviations.

\subsubsection{Machine-transcription parsing}

ATCO2-PL-set provides machine transcription results in word-confusion
network format. Each row is treated as one decoding step and contains
the source path, speaker label, start time, duration, candidate token
list, and token confidence scores. For each step, the candidate token
with the highest confidence is selected as the decoded token. Let
$W_i$ denote the candidate token set at step $i$, and let $P(w)$ denote
the confidence score of token $w$. The selected token is defined as
\begin{equation}
    \mathrm{Text}_i = \arg\max_{w \in W_i} P(w).
\end{equation}

The parsed intermediate record contains the source transcription file
identifier, start time, end time, speaker label, candidate call-sign
field, and decoded text. These records provide the basic units for
speaker-turn segmentation and subsequent call-sign matching.

\subsubsection{Speaker-turn segmentation}

Speaker turns are segmented according to the speaker labels in the
machine-transcription records. Consecutive decoding steps with the same
speaker label are assigned to the same speaker turn. Once the speaker
label changes, the current token buffer is flushed and concatenated into
one utterance. The utterance start time is taken from the first step in
the turn, and the utterance end time is taken from the last step in the
turn. No additional VAD-based or silence-threshold-based segmentation is
applied after transcript parsing.

For the $j$-th speaker turn, the utterance text is written as
\begin{equation}
    S_j =
    \mathrm{Concat}
    \left(
    \{\mathrm{Text}_i \mid \delta(\mathrm{Speaker}_i,j)\}
    \right),
\end{equation}
where $\delta(\mathrm{Speaker}_i,j)$ indicates whether the $i$-th
decoding step belongs to the $j$-th speaker turn.

\subsubsection{Filtering}

For ATCO2-PL train/validation, a valid call sign is determined by both
metadata and transcript evidence. First, the call sign must be included
in the normalized candidate dictionary built from ADS-B-derived
flight-aid metadata and supplementary call-sign entries. Second, the
accepted segment must contain utterance-level call-sign evidence. The
decoded utterance must match one of the normalized spoken full call
signs, or its candidate call-sign field must be resolved through the
valid dictionary. Thus, the metadata provides the candidate set, while
the decoded transcript provides the final utterance-level evidence.

SNR is read from the original ATCO2 organization rather than
re-estimated from the waveform. Specifically, the source audio path
encodes an integer SNR bin. If the SNR bin cannot be parsed, or if the
parsed value is not greater than 10, the segment is discarded from the
ATCO2-PL train/validation export. For ATCO2-PL train/validation, a
segment is retained if
\begin{equation}
\label{eq:clean}
D_{\mathrm{filtered}} =
\{d_i \mid
\mathrm{SNR}_i > 10,\ 
\mathrm{Dur}_i \geq 1.0,\ 
\mathrm{CallSign}_i \in C_{\mathrm{valid}}
\}.
\end{equation}
Here, $\mathrm{Dur}_i$ is computed from the speaker-turn start and end
times, and $C_{\mathrm{valid}}$ denotes the normalized valid call-sign
dictionary. Retained audio is cut from the original waveform and
exported as FLAC. For the finalized ATCO2-PL train/validation export,
201.26 h are retained after filtering.

The held-out test subset is processed separately. The ATCO2-test-set
provides complete human-annotated audio clips paired with transcripts
rather than speaker-turn segments generated from machine transcription.
Therefore, test examples are exported as complete human-annotated clips
whose transcripts contain normalized full call signs. The ATCO2-PL
duration threshold, $\mathrm{Dur}_i \geq 1.0$ s, is applied only to
train/validation. This explains why the test split can contain clips
shorter than 1.0 s.

\subsection{LLM-Assisted Call-Sign Caption Generation}
\label{subsec:llm_augmentation}

After preprocessing, each retained utterance is enriched with textual
descriptions. The initial JSON metadata contains a \texttt{text} list
with two types of descriptions. The first is a call-sign description,
which explicitly states that a specific aircraft call sign appears in
the recording. The second is a transcript description, which stores the
text content of the recording. LLM augmentation appends additional
ATC-style captions to the same \texttt{text} list. This stage builds on
the transformer-based language-modeling line of work and uses DeepSeek
as the generation backend~\cite{vaswani2017attention,devlin2019bert,liu2024deepseek}.

Given a transcript $T_i$ and a normalized call sign $\tilde{c}_i$, the
LLM is prompted to generate a concise ATC-style caption:
\begin{equation}
    y_i = \mathcal{M}_{\mathrm{LLM}}(T_i, \tilde{c}_i),
\end{equation}
where $\mathcal{M}_{\mathrm{LLM}}$ denotes the LLM used for caption
generation. The generated caption is expected to preserve the target
aircraft identity and summarize the operational content of the
utterance. When a potential mismatch between the transcript and the
normalized call-sign description is detected, a call-sign correction
record is retained for later quality analysis rather than silently
overwriting the original transcript.

Table~\ref{tab:stats} summarizes the augmentation outcome on the
training split. The number of captions increases from 266,042 to
640,460, while the number of unique captions rises from 119,435 to
475,047. In addition, 3,281 generated captions contain explicit
call-sign correction records, indicating that the correction stage
remains useful after rule-based filtering.

\begin{table}[t]
\centering
\caption{ATCCaps: augmentation effects of LLM enrichment on the training split.}
\label{tab:stats}
\footnotesize
\begin{tabularx}{\columnwidth}{Yccc}
\toprule
\textbf{Metric} & \textbf{Original} & \textbf{Augmented} & \textbf{Change} \\
\midrule
\#Captions & 266,042 & 640,460 & $2.41\times$ \\
\#Unique captions & 119,435 & 475,047 & $3.98\times$ \\
\#Captions with call-sign correction & 0 & 3,281 & +3,281 \\
\bottomrule
\end{tabularx}
\end{table}

Table~\ref{tab:ATCCapsexamples} presents examples from the unaugmented
training set, the augmented training set, and the test set.

\begin{table}[t]
\centering
\caption{Examples from ATCCaps subsets.}
\label{tab:ATCCapsexamples}
\footnotesize
\begin{tabularx}{\columnwidth}{c c Y}
\toprule
\textbf{Subset} & \textbf{Type} & \textbf{Example} \\
\midrule
\multirow{2}{*}{Train}
& Transcript & ``Oscar Kilo Papa Mike Bravo descend flight level one hundred.'' \\
& Call-sign desc. & ``The aircraft call sign \emph{Oscar Kilo Papa Mike Bravo} appears in this recording.'' \\
\midrule
\multirow{3}{*}{Train + Aug.}
& Transcript & same as above \\
& Call-sign desc. & same as above \\
& Extra desc. & ``Aircraft \emph{Oscar Kilo Papa Mike Bravo} is instructed to descend to flight level one hundred.'' \\
\midrule
\multirow{2}{*}{Test}
& Human transcript & ``Lufthansa eight five heavy, cleared for takeoff runway two five right.'' \\
& Call-sign desc. & ``The aircraft call sign \emph{Lufthansa eight five heavy} appears in this recording.'' \\
\bottomrule
\end{tabularx}
\end{table}

\subsection{Dataset Export Format}
\label{subsec:export_format}

The retained audio clips are exported in FLAC format. For the training
and validation splits, FLAC files and JSON files use the same sequential
numeric stem. For the ATCO2-test-set export, the original test-audio
filename information is preserved to maintain traceability.

Each JSON file contains a core \texttt{text} field:
\begin{quote}
\footnotesize
\ttfamily
\{\\
\quad "text": [\\
\quad\quad "The aircraft call sign '...' appears in this recording.",\\
\quad\quad "The text content of this recording is: '...'.",\\
\quad\quad "..."\\
\quad ]\\
\}
\end{quote}
After LLM augmentation, generated ATC-style captions are appended to the
same \texttt{text} list. Information such as SNR, duration, source month,
and filtering decisions is retained in construction logs rather than
being separately stored as structured fields in the current JSON
metadata. This export format provides multiple textual descriptions for
each audio segment and supports audio-text retrieval, call-sign-aware
caption matching, and ASR-related evaluation.

\section{Dataset Analysis and Quality Validation}
\label{sec:analysis}

This section characterizes ATCCaps from the perspective of dataset
scale, call-sign coverage, filtering reliability, and caption quality.
The goal is to verify that the retained corpus provides meaningful
call-sign-aware supervision while preserving realistic ATC speech
conditions. The analysis focuses on the dataset itself rather than on
model-specific improvements.

\subsection{Overall Statistics and Split Characteristics}
\label{subsec:overall_statistics}

Table~\ref{tab:split_stats} reports the final split-level statistics of
ATCCaps. The training and validation splits are derived from
ATCO2-PL-set, while the evaluation subset is filtered from the
human-annotated ATCO2-test-set provided by ATCO2. In total, ATCCaps
contains 202.94 hours of curated ATC audio, 170,385 utterances, and
922 unique normalized call signs.

The training split contains 160.44 hours of audio and 133,021
utterances, while the validation split contains 40.82 hours and 35,894
utterances. Both splits are produced from ATCO2-PL-set after
speaker-turn parsing and filtering. The evaluation subset contains
1.68 hours and 1,470 utterances from ATCO2-test-set. Since ATCO2-test-set
provides complete human-annotated clips rather than \texttt{.cnet}-based
speaker-turn segments, the ATCO2-PL duration threshold is not applied to
this subset. This is why the minimum duration in the evaluation subset is
0.46 s, although the ATCO2-PL train/validation filtering rule requires
duration $\geq 1.0$ s.

\begin{table}[!t]
\centering
\caption{ATCCaps split statistics and call-sign overlap.}
\label{tab:split_stats}
\label{tab:callsign_overlap}
\scriptsize
\setlength{\tabcolsep}{3pt}
\begin{minipage}{\columnwidth}
\centering
\textbf{(a) Split statistics}\\[2pt]
\resizebox{\columnwidth}{!}{%
\begin{tabular}{llcccccc}
\toprule
\textbf{Split} & \textbf{Source} & \textbf{Hours} & \textbf{\#Utt.} & \textbf{\#CS} & \textbf{Avg (s)} & \textbf{Range (s)} & \textbf{Role} \\
\midrule
Train & ATCO2-PL-set & 160.44 & 133,021 & 733 & 4.34 & 1.00--38.48 & Training \\
Val & ATCO2-PL-set & 40.82 & 35,894 & 389 & 4.09 & 1.00--34.44 & Model selection \\
Eval. & ATCO2-test-set & 1.68 & 1,470 & 366 & 4.11 & 0.46--23.67 & Final evaluation \\
\bottomrule
\end{tabular}
}
\vspace{1pt}
\emph{Note:} the duration filter of $\geq 1.0$ s is applied only to
ATCO2-PL train/validation segments. The evaluation subset is filtered
from complete human-annotated ATCO2-test-set clips.
\vspace{4pt}

\textbf{(b) Call-sign overlap}\\[2pt]
\begin{tabularx}{0.95\columnwidth}{Yccc}
\toprule
\textbf{Split Pair} & \textbf{\#Overlap CS} & \textbf{Overlap (\%)} & \textbf{\#Unseen} \\
\midrule
Train--Val & 330 & 84.83 & 59 \\
Train--Eval. & 233 & 63.66 & 133 \\
Val--Eval. & 139 & 37.98 & 227 \\
\bottomrule
\end{tabularx}
\end{minipage}
\end{table}

\subsection{Call-Sign Coverage and Cross-Split Overlap}
\label{subsec:callsign_coverage}

The call-sign statistics in Table~\ref{tab:callsign_overlap}(b) show
that ATCCaps supports both seen and unseen call-sign evaluation. The
training and evaluation subsets share 233 call signs, which provides
seen-call-sign cases for stable evaluation. At the same time, the
evaluation subset contains 133 call signs that are not observed in the
training split. This setting creates an open-set call-sign generalization
condition and is useful for evaluating call-sign recognition,
call-sign-aware retrieval, and audio-text grounding under realistic
deployment assumptions.

The validation and evaluation subsets have a lower overlap of
139 call signs, leaving 227 validation-unseen call signs in the
evaluation subset. This indicates that the ATCO2-test-set-derived
evaluation subset does not simply reproduce the validation distribution.
The resulting split design therefore preserves both shared call-sign
patterns and unseen call-sign cases, which is important for assessing
whether a model can generalize beyond memorized aircraft identifiers.

\subsection{Filtering Effectiveness and Rule-Consistency Audit}
\label{subsec:filtering_validation}

The ATCO2-PL source material is large and acoustically diverse, but it
also contains noisy pseudo transcripts, low-SNR recordings, short
segments, malformed transcriptions, and utterances without valid
call-sign evidence. The filtering stage is therefore designed to retain
call-sign-bearing speech segments with usable acoustic and textual
quality. Table~\ref{tab:filtering_diagnostics} summarizes month-wise
retention diagnostics after filtering.

\begin{table}[t]
\centering
\caption{Month-wise retention diagnostics after ATCCaps filtering.}
\label{tab:filtering_diagnostics}
\small
\setlength{\tabcolsep}{2pt}
\begin{tabular}{lccccc}
\toprule
\textbf{Month} & \textbf{Raw (h)} & \textbf{Filt. (h)} & \textbf{Ret. (\%)} & \textbf{\#CS} & \textbf{\#Seg.} \\
\midrule
2020\_10 & 88.2 & 2.35 & 2.66 & 76 & 1,782 \\
2020\_11 & 153.9 & 10.28 & 6.68 & 107 & 7,705 \\
2021\_01 & 116.1 & 4.50 & 3.88 & 107 & 3,385 \\
2021\_02 & 82.3 & 4.08 & 4.96 & 120 & 3,063 \\
2021\_03 & 70.3 & 3.85 & 5.48 & 103 & 3,150 \\
2021\_04 & 166.4 & 9.10 & 5.47 & 300 & 7,952 \\
2021\_05 & 340.2 & 16.37 & 4.81 & 767 & 15,713 \\
2021\_06 & 443.3 & 20.26 & 4.57 & 794 & 19,657 \\
2021\_07 & 486.5 & 18.98 & 3.90 & 406 & 18,295 \\
2021\_08 & 576.5 & 12.86 & 2.23 & 401 & 11,983 \\
2021\_09 & 553.0 & 16.39 & 2.96 & 366 & 15,016 \\
2021\_10 & 675.9 & 38.60 & 5.71 & 498 & 34,967 \\
2021\_11 & 422.5 & 20.76 & 4.91 & 408 & 18,643 \\
2021\_12 & 432.4 & 20.83 & 4.82 & 441 & 19,471 \\
2022\_01 & 402.0 & 14.65 & 3.64 & 378 & 13,493 \\
\bottomrule
\end{tabular}
\end{table}

The retention rate ranges from 2.23\% to 6.68\% across months, showing
that the filtering process is highly selective. This selectivity is
expected because a retained ATCO2-PL segment must satisfy three
conditions simultaneously: sufficient SNR, sufficient duration, and
valid call-sign evidence in the transcript or candidate field. The
variation across months also reflects the heterogeneity of the original
ATCO2-PL recordings in channel condition, transcript quality, and
call-sign availability. The finalized train/validation export contains
201.26 hours, 168,915 segments, and 792 unique call signs after filtering
and consolidation.

To check whether the rejection rules remove samples for appropriate
reasons, we conduct a rule-consistency audit on removed samples. For
each rejection category, 25 removed samples are inspected and labeled as
reasonable or questionable. This audit is intended as a consistency check
of the filtering rules rather than a large-scale human annotation study.
The results are shown in Table~\ref{tab:rule_audit}.

\begin{table}[t]
\centering
\caption{Rule-consistency audit of removed samples.}
\label{tab:rule_audit}
\footnotesize
\setlength{\tabcolsep}{3pt}
\resizebox{\columnwidth}{!}{%
\begin{tabular}{lcccc}
\toprule
\textbf{Filter Reason} & \textbf{\#Sampled} & \textbf{\#Reasonable} & \textbf{\#Questionable} & \textbf{Precision (\%)} \\
\midrule
Invalid call sign & 25 & 23 & 2 & 92.0 \\
Low SNR & 25 & 25 & 0 & 100.0 \\
Short duration & 25 & 24 & 1 & 96.0 \\
Malformed transcript & 25 & 24 & 1 & 96.0 \\
\bottomrule
\end{tabular}
}
\end{table}

The audit shows that low-SNR removals are fully consistent with manual
inspection in the sampled set. Invalid-call-sign, short-duration, and
malformed-transcript removals each achieve at least 92\% precision. The
few questionable cases mainly arise from borderline transcripts or
ambiguous call-sign evidence. Overall, the audit indicates that the
filtering rules remove large amounts of noisy or weakly grounded data
without relying on arbitrary pruning.

\subsection{Caption Diversity and Consistency}
\label{subsec:caption_quality}

LLM augmentation is used to enrich the textual supervision associated
with each retained training audio clip. The analysis in this subsection
examines both the benefit and the risk of this augmentation. The benefit
is measured through caption diversity and ATC keyword coverage, while
the risk is examined through call-sign mention rate, numeric-token
preservation, verbatim fallback rate, and qualitative inspection. These
metrics are important because call signs and numeric expressions are
safety-critical in ATC communication.

Table~\ref{tab:caption_quality}(a) reports automatic diversity metrics.
The number of captions increases from 266,042 to 640,460 after
augmentation, while the number of unique captions increases from
119,435 to 475,047. The vocabulary size also expands from 5,429 to
16,656. Distinct-1 and Distinct-2 improve after augmentation, indicating
broader lexical and phrasal coverage. The Self-BLEU score also increases,
which suggests that many generated descriptions remain close
paraphrases of the same ATC event. This behavior is reasonable for ATC
speech because the phraseology is constrained and many utterances share
similar operational structures.

\begin{table}[t]
\centering
\caption{Caption diversity, consistency, and qualitative audit.}
\label{tab:caption_quality}
\footnotesize
\setlength{\tabcolsep}{3pt}
\textbf{(a) Automatic caption-diversity comparison.}\\[-1pt]
\resizebox{\columnwidth}{!}{%
\begin{tabular}{lccccccc}
\toprule
\textbf{Version} & \textbf{\#Captions} & \textbf{\#Unique} & \textbf{Avg Tok.} & \textbf{Vocab.} & \textbf{Dist.-1} & \textbf{Dist.-2} & \textbf{Self-BLEU $\downarrow$} \\
\midrule
Original & 266,042 & 119,435 & 15.98 & 5,429 & 0.0013 & 0.0258 & 0.3093 \\
Augmented & 640,460 & 475,047 & 12.78 & 16,656 & 0.0020 & 0.0326 & 0.5740 \\
\bottomrule
\end{tabular}
}
\vspace{3pt}

\textbf{(b) Consistency and task relevance.}\\[-1pt]
\resizebox{\columnwidth}{!}{%
\begin{tabular}{lcccc}
\toprule
\textbf{Version} & \textbf{CS Mention (\%)} & \textbf{ATC Keyword (\%)} & \textbf{Numeric (\%)} & \textbf{Verbatim (\%)} \\
\midrule
Original & 100.00 & 53.46 & 100.00 & 0.00 \\
Augmented & 71.34 & 65.68 & 51.59 & 13.26 \\
\bottomrule
\end{tabular}
}
\vspace{3pt}

\textbf{(c) Qualitative audit of 25 samples.}\\[-1pt]
\resizebox{\columnwidth}{!}{%
\begin{tabular}{lcccc}
\toprule
\textbf{Sample Set} & \textbf{CS Correct (\%)} & \textbf{Faithful (\%)} & \textbf{ATC-Natural (\%)} & \textbf{Informative (\%)} \\
\midrule
Augmented (25 samples) & 96.0 & 92.0 & 84.0 & 72.0 \\
\bottomrule
\end{tabular}
}
\end{table}

Table~\ref{tab:caption_quality}(b) reports consistency and task-relevance
metrics. ATC keyword coverage increases from 53.46\% to 65.68\%, showing
that the generated captions more often expose operational concepts such
as climb, runway, taxi, and frequency change. At the same time, the
call-sign mention rate of augmented captions is 71.34\%, and
numeric-token preservation is 51.59\%. These results show that LLM
augmentation increases textual variety and operational phrasing, while
explicit consistency checks remain necessary for call signs and numeric
tokens.

Table~\ref{tab:caption_quality}(c) gives a small qualitative audit of
25 augmented samples. Most inspected captions preserve the correct call
sign and remain faithful to the transcript, with 96.0\% call-sign
correctness and 92.0\% faithfulness. However, the ATC-naturalness and
informativeness scores are lower, indicating that some generated
captions are generic, terse, or insufficiently informative. These
findings support the use of augmented captions as caption-level
supervision for retrieval and grounding, while transcripts and explicit
call-sign labels remain the primary references for safety-critical
recognition tasks.

The above analysis characterizes ATCCaps as a curated call-sign-aware
audio-text resource. The corpus preserves realistic ATC acoustic
conditions from ATCO2, adds normalized call-sign supervision, and
provides caption-level textual descriptions for audio-text grounding.
The split statistics show that the training and validation subsets
provide large-scale call-sign-bearing speech, while the ATCO2-test-set-
derived evaluation subset provides a human-annotated reference for final
evaluation. The call-sign overlap analysis supports both seen and unseen
call-sign evaluation. The filtering audit indicates that the retained
data are selected through consistent quality-control rules. Finally, the
caption analysis shows that LLM augmentation expands textual supervision
but requires explicit attention to call-sign and numeric fidelity.

These dataset properties motivate the benchmark protocols in the next
section, where ATCCaps is evaluated for ASR, call-sign matching, and
call-sign-aware audio-text retrieval.

\section{Reference Evaluation}
\label{sec:benchmarks}

This section provides reference evaluations for ATCCaps. The purpose is
to show how the dataset can be used for representative ATC speech tasks,
rather than to introduce a new model. We consider three tasks that match
the design of ATCCaps: ASR evaluation, call-sign matching, and
call-sign-aware audio-text retrieval.

\subsection{Evaluation Setup}
\label{subsec:evaluation_setup}

\textbf{Evaluation data.}
For in-domain evaluation, we use the ATCCaps evaluation subset derived
from the human-annotated ATCO2-test-set. This subset contains
call-sign-bearing utterances with human transcripts. For external
cross-corpus reference, we use the publicly available \texttt{valid}
split of UWB-ATCC. Since the available UWB-ATCC release in our
environment provides a validation split rather than a unified official
test split, results on UWB-ATCC are used only as cross-corpus reference
points.

\textbf{Training variants.}
For audio-text tasks, we compare two training variants. \emph{ATCCaps}
uses the original text list consisting of call-sign descriptions and
transcript descriptions. \emph{ATCCaps\textsuperscript{+}} additionally
uses LLM-generated ATC-style captions. This comparison evaluates whether
caption-level augmentation changes audio-text grounding and call-sign
matching performance.

\textbf{Reference models.}
For ASR, we evaluate an ATC-finetuned Whisper Large v3 checkpoint,
\path{jacktol/whisper-large-v3-finetuned-for-ATC}, without additional
fine-tuning or language-model rescoring. For audio-text retrieval and
call-sign matching, we use CLAP~\cite{elizalde2023clap} as a
representative contrastive audio-text baseline. This choice follows the
broader use of contrastive and attentional feature-fusion objectives for
cross-modal representation learning~\cite{faghri2017improving,dai2021attentional,ye2022cross}.
We intentionally keep
the main evaluation model-agnostic and avoid introducing additional
model-specific components in the main text.

\textbf{Metrics.}
For ASR, we report word error rate (WER) and character error rate (CER).
Reference transcripts are normalized by lowercasing, removing fixed
metadata prefixes, stripping non-alphanumeric symbols except apostrophes,
and collapsing repeated whitespace. For retrieval, we report Recall@1,
Recall@5, Recall@10, and mAP@10 in both audio-to-text
(A$\rightarrow$T) and text-to-audio (T$\rightarrow$A) directions. For
call-sign matching, we report accuracy (ACC), precision, recall, F1, and
AUC.

\subsection{ASR Evaluation}
\label{subsec:asr_evaluation}

ASR evaluation verifies whether the ATCCaps evaluation subset can serve
as a realistic ATC-ASR test resource. The ATC-finetuned Whisper baseline
is applied directly to each audio clip, and the predicted transcript is
compared with the human transcript in the ATCCaps evaluation subset. The
same inference procedure is applied to UWB-ATCC valid for external
reference.

\begin{table}[t]
\centering
\caption{ASR reference results on released ATC evaluation splits.}
\label{tab:asr_baseline}
\footnotesize
\setlength{\tabcolsep}{3pt}
\begin{tabular}{lccc}
\toprule
\textbf{Evaluation set} & \textbf{Setting} & \textbf{WER} & \textbf{CER} \\
\midrule
ATCCaps eval. & ATC zero-shot & 0.1485 & 0.0944 \\
UWB-ATCC valid & ATC zero-shot & 0.2002 & 0.1288 \\
\bottomrule
\end{tabular}
\end{table}

Table~\ref{tab:asr_baseline} shows that the ATCCaps evaluation subset is
usable for ASR evaluation with an ATC-adapted recognizer. The baseline
achieves 0.1485 WER and 0.0944 CER on ATCCaps evaluation data. On
UWB-ATCC valid, the same model obtains 0.2002 WER and 0.1288 CER. The
gap reflects cross-corpus differences in channel condition, recording
style, and domain coverage. Qualitative inspection indicates that
remaining errors often involve long call signs, location names, and
dense ATC phraseology, which supports the need for call-sign-aware
evaluation beyond aggregate WER and CER.

\begin{table}[H]
\centering
\caption{Call-sign matching performance using CLAP.}
\label{tab:callsign_matching}
\scriptsize
\setlength{\abovecaptionskip}{1pt}
\setlength{\belowcaptionskip}{1pt}
\renewcommand{\arraystretch}{0.82}
\setlength{\tabcolsep}{2.5pt}
\resizebox{\columnwidth}{!}{%
\begin{tabular}{llccccc}
\toprule
\textbf{Eval. set} & \textbf{Pretrain} & \textbf{ACC} & \textbf{Prec.} & \textbf{Rec.} & \textbf{F1} & \textbf{AUC} \\
\midrule
ATCCaps eval. & ATCCaps & 0.8854 & 0.8634 & 0.9156 & 0.8887 & 0.9483 \\
ATCCaps eval. & ATCCaps\textsuperscript{+} & \textbf{0.8912} & \textbf{0.8705} & \textbf{0.9190} & \textbf{0.8941} & \textbf{0.9501} \\
UWB-ATCC & ATCCaps & \textbf{0.7677} & \textbf{0.7708} & 0.7619 & 0.7663 & \textbf{0.8460} \\
UWB-ATCC & ATCCaps\textsuperscript{+} & 0.7600 & 0.7393 & \textbf{0.8033} & \textbf{0.7700} & 0.8359 \\
\bottomrule
\end{tabular}
}
\end{table}

\subsection{Call-Sign Matching}
\label{subsec:callsign_matching}

Call-sign matching directly evaluates the core supervision provided by
ATCCaps. Given an audio segment and a call-sign-related text description,
the model predicts whether the pair matches. Positive pairs are formed
from the matched audio and its call-sign description. Negative pairs are
formed by pairing the audio with mismatched call-sign descriptions from
the evaluation pool. The task is evaluated as binary matching.

Table~\ref{tab:callsign_matching} shows that CLAP trained on ATCCaps can
perform call-sign matching with high AUC on the ATCCaps evaluation
subset. Caption augmentation provides small improvements on the
in-domain evaluation subset, increasing ACC from 0.8854 to 0.8912 and
F1 from 0.8887 to 0.8941. On UWB-ATCC, performance decreases because of
cross-corpus domain shift. The caption-augmented variant improves recall
and F1 but slightly reduces precision and AUC, indicating a trade-off
between broader matching coverage and stricter discrimination under
domain shift.

\begin{table*}[t]
\centering
\caption{Call-sign-aware audio-text retrieval performance using CLAP.}
\label{tab:retrieval}
\scriptsize
\setlength{\tabcolsep}{3pt}
\begin{tabular}{llcccccccc}
\toprule
\textbf{Pretrain} & \textbf{Eval. set} &
\multicolumn{4}{c}{\textbf{A$\rightarrow$T}} &
\multicolumn{4}{c}{\textbf{T$\rightarrow$A}} \\
\cmidrule(lr){3-6}\cmidrule(lr){7-10}
& & \textbf{R@1} & \textbf{R@5} & \textbf{R@10} & \textbf{mAP@10}
& \textbf{R@1} & \textbf{R@5} & \textbf{R@10} & \textbf{mAP@10} \\
\midrule
ATCCaps & ATCCaps eval. & \textbf{0.3687} & \textbf{0.6136} & \textbf{0.7122} & \textbf{0.4760} & \textbf{0.3880} & \textbf{0.5628} & 0.6257 & \textbf{0.4583} \\
ATCCaps\textsuperscript{+} & ATCCaps eval. & 0.3252 & 0.5803 & 0.6864 & 0.4360 & 0.3169 & 0.5410 & \textbf{0.6257} & 0.4089 \\
ATCCaps & UWB-ATCC & \textbf{0.0713} & 0.1903 & 0.2864 & \textbf{0.1275} & 0.1169 & 0.2922 & 0.4069 & 0.1945 \\
ATCCaps\textsuperscript{+} & UWB-ATCC & 0.0522 & \textbf{0.1961} & \textbf{0.2979} & 0.1149 & \textbf{0.1364} & \textbf{0.3225} & \textbf{0.4481} & \textbf{0.2195} \\
\bottomrule
\end{tabular}
\end{table*}

\subsection{Call-Sign-Aware Audio-Text Retrieval}
\label{subsec:retrieval}

Audio-text retrieval evaluates whether ATC audio can be grounded to the
corresponding textual description. In the A$\rightarrow$T direction, an
audio clip is used as the query and the model retrieves the paired text
from the candidate pool. In the T$\rightarrow$A direction, the text is
used as the query and the model retrieves the paired audio. We evaluate
CLAP pretrained on ATCCaps and ATCCaps\textsuperscript{+}.

Table~\ref{tab:retrieval} reports retrieval results in both directions.
On the ATCCaps evaluation subset, the original ATCCaps training variant
performs better than the caption-augmented variant for most CLAP
retrieval metrics. This indicates that LLM-generated captions do not
automatically improve in-domain retrieval, especially when the original
call-sign and transcript descriptions already provide strong alignment
signals. On UWB-ATCC, caption augmentation improves several
cross-domain T$\rightarrow$A metrics and high-rank A$\rightarrow$T
recall, suggesting that additional textual variation may help under
domain shift. These mixed results are consistent with the caption
quality analysis in Section~\ref{subsec:caption_quality}: LLM captions
increase phraseological diversity, but their call-sign and numeric
fidelity must be evaluated separately.

\section{Conclusion}
\label{sec:conclusion}

This paper presents ATCCaps, a call-sign-aware ATC speech dataset with
caption-level audio-text supervision. Built from real ATC radiotelephony
recordings, ATCCaps contains 202.94 hours of curated audio, 170,385
utterances, and 922 unique normalized call signs. The construction
pipeline combines confidence-aware transcript parsing, ADS-B-derived
call-sign metadata, call-sign normalization, quality filtering, and
LLM-assisted caption generation, providing paired audio, transcripts,
call-sign descriptions, and ATC-style captions for call-sign-aware ASR,
call-sign matching, and audio-text grounding.

Dataset analyses show that ATCCaps provides both scale and call-sign
diversity, including seen and unseen call-sign cases in the evaluation
subset derived from the human-annotated ATCO2-test-set. Filtering audits
support the consistency of the preprocessing rules, while caption
analysis shows that LLM augmentation increases textual diversity and
ATC-related phrasing. Reference evaluations further demonstrate that
ATCCaps can support ASR evaluation, call-sign matching, and
call-sign-aware audio-text retrieval. Future work will extend ATCCaps
with larger human-verified caption audits, structured metadata fields,
and entity-level ASR metrics such as call-sign accuracy and numeric
accuracy.
\bibliographystyle{IEEEtran}
\bibliography{refs}

\end{document}